# Temperature Dependence Analysis of the NIR Spectra of Liquid Water confirms the existence of two phases, one of which is in a coherent state

**Paolo Renati** [1,†,‡], **Zoltan Kovacs** [2], **Antonella De Ninno** [3] **and Roumiana Tsenkova** [4]

[1] (†) University of Catania, Faculty of Mathematical, Physical and Natural Sciences, Department of Physics and Astronomy, LNS-INFN, Via S. Sofia 64, I-95123 Catania, Italy; (‡) Atena s.r.l., Via Marzeno 65, 48013, Brisighella (RA), Italy; paolo.renati@gmail.com ;

[2] Szent István University, Faculty of Food Science, Department of Physics and Control, 14-16 Somlói str., Budapest 1118, Hungary; zoltan.kovacs2.unicorvinus@gmail.com ;

[3] ENEA, FSN-TECFIS_DIM, Research Centre Frascati, via Enrico Fermi 45 00044 Frascati (Roma) Italy; antonella.deninno@enea.it ;

[4] Kobe University, Graduate School of Agricultural Science, Biomeasurement Technology Laboratory, 1-1 Rokkodai, Nada-ku, Kobe 657-8501, Japan; rtsen@kobe-u.ac.jp ;

**\*** Correspondence: paolo.renati@gmail.com ; (P.R.)

**Abstract: Background**: isosbestic (equal absorption) points in the IR and NIR spectra of liquid water are a well-known feature and they witness the existence of two populations of oscillators in the probed system. Despite it is a well-known experimental fact, in the mainstream molecular dynamics approach the proposed theoretical explanations for it are not able to elucidate which is the physical reason *why* such a "cut-off frequency" (at the isosbestic point) does exist. **Methods**: We investigate pure Milli-Q water on increasing the temperature in the vis-NIR range (400-2500 nm). We specifically payed attention to the first overtone region (1300-1600 nm) of the OH-bond stretching-mode where an isosbestic point has been observed. **Results**: A second derivative analysis clearly shows two modes, which can be assigned to water molecules involved in different "hydrogen bonding" configurations whose relative abundance is controlled by the temperature. We have also observed that the ratio of these modes follows a van't Hoff behavior supporting that their energy difference (energy gap) is independent on the temperature. Furthermore, a log-log plot shows a scale invariance of the population ratio with respect to the perturbation (temperature), confirming the existence of a long-range correlated dynamics in the liquid. **Conclusions**: We show that the two phases differences between energy and entropy estimated from the experimental data can be compared with the prediction of Quantum Electro Dynamics (QED) showing a remarkable agreement.

## 1. Introduction

Nowadays, water molecular structures have been widely investigated with vibrational infrared (IR) and the respective visible-near infrared (vis-NIR) overtone spectroscopy. Water is a strong absorber of IR energy and, as a result, this method is only suited to analyse small or very thin samples. In contrast, water absorption in the vis-NIR range is several times weaker, meaning that this spectral range can be used to analyse thicker samples. By using the vis-NIR region, it is very easy to acquire spectral data for monitoring and measuring aqueous systems (also biological ones). A spectrum of water in the vis-NIR range, 400–2500 nm, presents three dominant broad peaks known as the first, second and third overtones and a few other distinct but smaller peaks, known as combination-band peaks [1].

In the literature, NIR [2-4] (and IR) spectroscopic measurements of liquid water were repeatedly performed in a wide range of temperature and gave rise to a lively debate among scientists. The Molecular Dynamics (MD) modelling of the FTIR and Raman spectra, could seem to support the picture of a continuum system of H bonded water molecules. However, in the last decade, many works have appeared in the literature concerning the experimental evidence of the coexistence of two phases (low-density - high-correlation and high-density - low-correlation). This fact has been recorded (both in super-cooled and normal, both bulk and confined) through spectroscopic techniques like X-rays Absorption [5], small angle X-rays scattering [6], time-resolved optical Kerr effect [7] and even in computational simulation [8] such a bi-phasicity has been modelled. MD calculations are mainly based on the mutual independence of the constituting molecules, however, it has been suggested, starting from first principles, that, above a density threshold and below a critical temperature, a collective state can be induced in the physical systems. Water represents a remarkable example of such a general principle [9-14].

In the following section we analyze some important features of the vis-NIR first overtone band such as the existence of an isosbestic point and the linearity of van't Hoff behavior and the linear dependence of absorbance with the logarithm of the temperature in vicinity of the isosbestic point.

## 2. Materials and Methods

The effect of temperature perturbation on the water near infrared spectra is a well-studied topic in the literature [2,3,15], therefore, experiments were performed on Milli-Q water in two different arrangements. The Milli-Q water was produced by Milli-Q purification system (Millipore, Molsheim, France, resistance = 18 MΩ). We probed Milli-Q water in the temperature range between 20 and 74°C. The spectral acquisition was performed on the full range by steps of 2 °C resulting in 28 temperature steps using an external thermalizing bath. The spectra were acquired after the careful temperature equilibration (<0.1°C) performing three consecutive reading of the same sample for each temperature and they were used to evaluate the error bars in figure 6 and table 1. Reference spectra were recorded before each sample. Sample spectrum, followed by reference spectrum, were acquired each 70 seconds. Such a time is necessary (1) for the instrument to acquire the whole spectrum, (2) to remove the sample from the beam path, (3) to acquire the reference spectrum and (4) to re-position the cuvette. Such handling is provided by the default working protocol of the instrumental setup. The temperature at which 55 respective spectra were acquired was monitored using thermocouple in contact with the cuvette. Temperature and humidity were kept constant in the room (16.6°C and 14% RH). We used rubber plug to close the cuvette and in addition to it we wrapped by Parafilm M® the top of the cuvette and the rubber plug together to reduce evaporation.

About spectral acquisition we used FOSS-XDS spectrometer (FOSS NIR-Systems, Inc., Hoganas, Sweden) equipped with Rapid Liquid Analyzer module including temperature-controlled cuvette holder was used to measure transmittance spectra of the Milli-Q samples, dropping 1 ml of liquid in the cuvette. Spectral acquisition was performed saving three consecutive spectra in the range of 400-2500 nm at 0.5 nm spectral step. All the saved spectra were a result of the average of 32 successive scans. In fig. 1 we show a detail of the sample-holder.

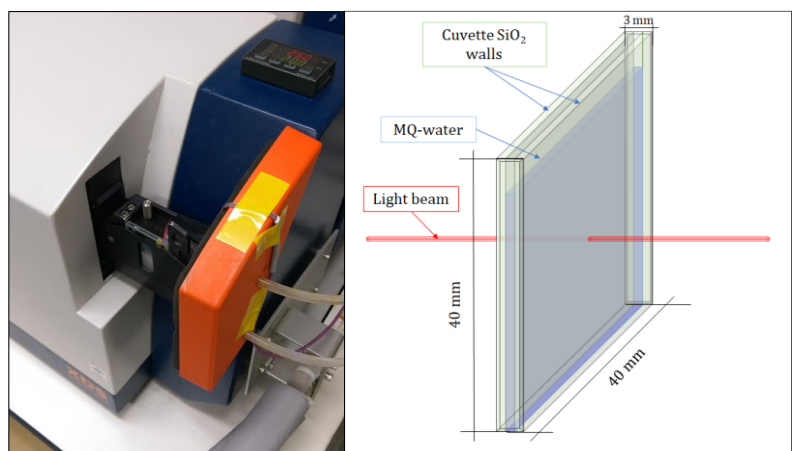

**Figure 1.** experimental layout showing the sample-holder (on the left) and a scheme of the transmittance measurement setting and the geometry of the cuvette (on the right).

The thermal bath with continuous water circulation was attached to the Rapid Liquid Analyzer module to ensure the required temperature of the sample during the scanning. Due to a certain amount of temperature-loss in the tubing of the thermal bath, the temperature of the thermal bath was set always as much hotter/colder as was necessary to reach the required temperature of the sample in the holder during the experiment (the system is an On-Off thermoregulation providing a precision < 0.1 °C). Before the spectral acquisition the samples were incubated for 90 seconds to equilibrate to the temperature (except for the Consecutive Experiment). FOSS-XDS instrument was operated by using VISION 3.5 software (FOSS NIR Systems, Inc., Hoganas, Sweden).

Changing the temperature of water results a change of baseline of the spectrum beside the wavelength shift. Mathematical baseline correction was applied to remove the effect of baseline. The steps of the baseline correction are introduced in the Appendix B (Figure B1 and Figure B2).

In Fig. 2 a detail of the transmittance experimental layout is shown. The cuvette walls are made of silicon dioxide and, unless it is fully siloxane-ended on the surface (which requires specific treatments/preparations), this material is a moderate-to-medium hydrophilic material [16].

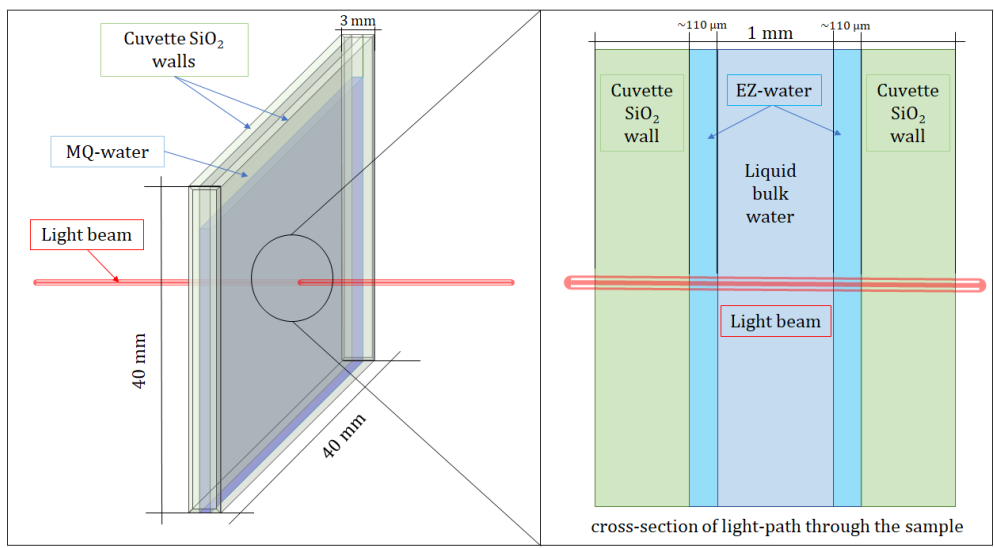

**Figure 2.** Sketch of the experimental transmittance layout: on the right side the schemed detail showing which is the physical situation experienced by water confined in standard quartz walls.

We highlight this aspect because it is important for a correct interpretation of our results, especially in comparison with a previous work [17], as will be explained in the discussion.

## 3. Experimental Data and Results

In fig.3 we show in different colors the spectra of Milli-Q water in the range 400-2500 nm for several temperatures, from 20° to 74° C, by 2°C steps (from blue to red and to yellow for low-to-high T spectra).

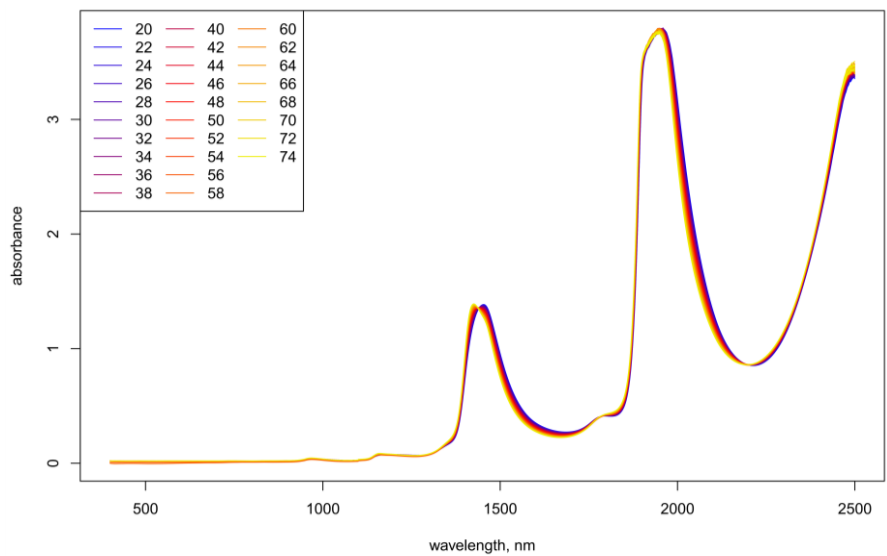

**Figure 3.** Raw absorbance spectra of Milli-Q water samples at 20-74 °C in the entire spectral range of 400-2500 nm. We can see the broad bands of the first overtone of OH (1300-1600 nm), the combination region of OH stretching and bending vibrations (1800-2100 nm).

In fig.4 the spectra in the first overtone range (1300-1600 nm) and their second derivatives are shown.

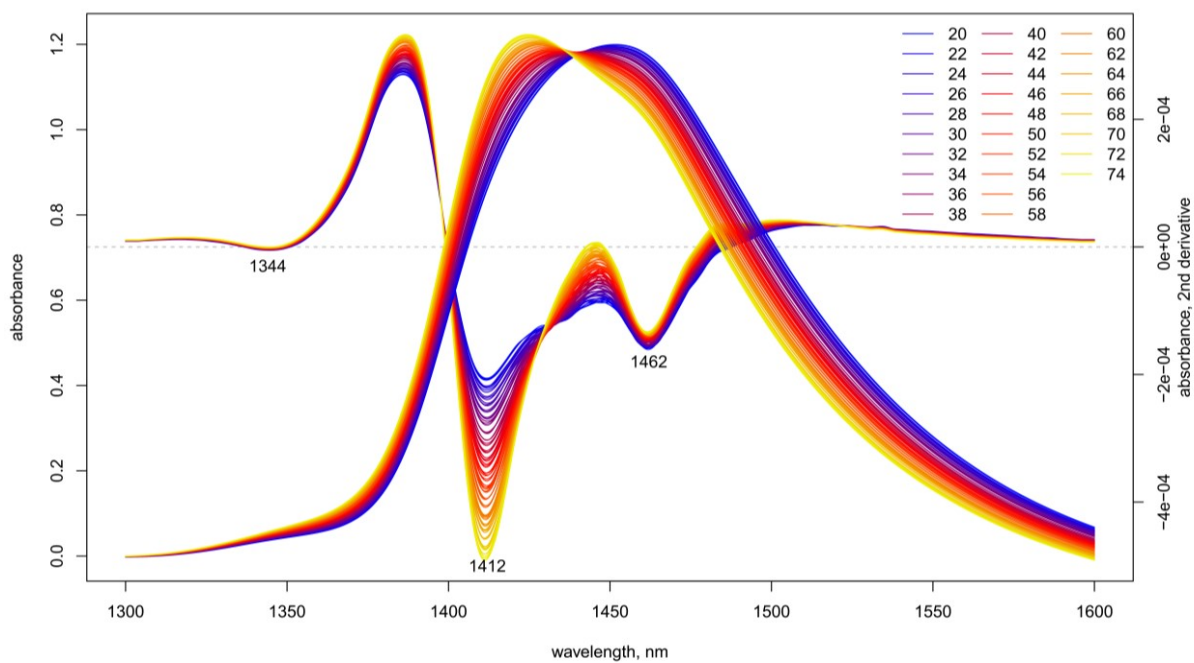

**Figure 4.** Baseline corrected and 2nd derivative (Savitzky-Golay filter using 2nd order polynomial and 17 points) absorbance spectra of Milli-Q water samples of 20-74 °C in the 1300-1600 nm range; note the two minima that reveal two kinds of oscillators around the isosbestic point

The isosbestic point is clearly visible around 1438 nm, peaked on the violet line. We identified it as the wavelength at which the optical density at the several tested temperatures shows the less variation (i.e.: its standard deviation is minimum). As it has been usually done for the IR range (3800-3000 cm-1), as in [17] for instance, to identify the components constituting the broad absorption peak we count the minima in the second derivative of the spectra in the first overtone range (see fig.4).

We can observe two deep minima showing that the minimum number of sub-bands contributing to the peak respectively at 1412 nm ($\lambda_2$) and 1462 nm ($\lambda_1$). Namely, the isosbestic point is the wavelength value $\lambda^*$, at which the spectra optical density, Abs*($\lambda$,T), taken at different temperatures exhibit the same absorption. Isosbestic points is the experimental manifestation of a general rule of sum involving two population of molecules represented by the spectral intensities $I_1$ and $I_2$ being $I_1+I_2$ = constant. It corresponds to the wavelength of equal absorption in a spectrum affected by a changing parameter. This means that the overall spectrum can be decomposed exactly into the sum of the two fractions, in the liquid system, having two different bond-lengths of the intramolecular O-H spring [17,18].

Different curves intersect in a crossing point. The crossing points are located along a curve $\lambda^*(T)$ defined by $\left.\frac{\partial Abs(\lambda,T)}{\partial T}\right|_{\lambda^*(T)} = 0$. If the curves intersect at a single (exact) point $\lambda^*$ does not depend on T at all. Actually, in our experimental setup there is a residual dependence on T, as can be seen in fig. 4: the isosbestic point is not an exact point but ranges in a region 8 nm wide. In Appendix A we show the detailed evaluation of the terms responsible for the deviation from the exact isosbestic case.

In fig. 5 we show an example (for T=36°C) of the spectral area divided at the isosbestic lambda 1438 nm and in table 1 we show the full list of the spectral area and their ratios and their standard deviation constituting the error bar in the van't Hoff plot shown in fig. 6.

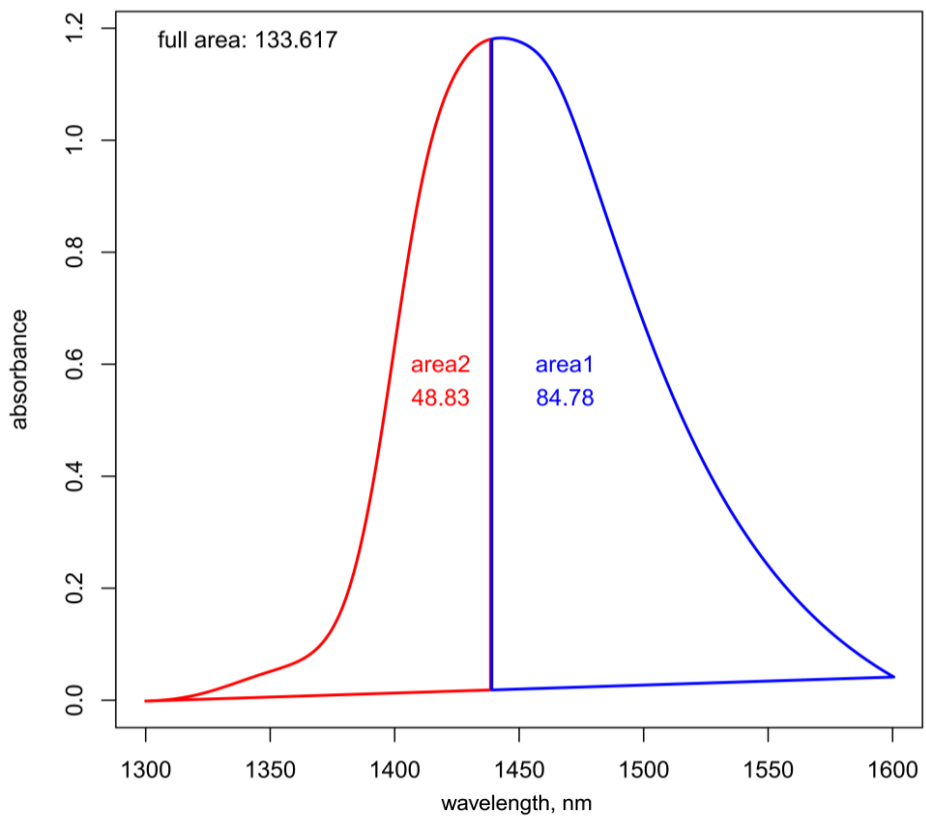

**Figure 5.** Example for T=36°C of the first overtone spectral intensity divided into the two components above and below the isosbestic point (at 1438 nm), representing the two components that, respectively, increases and decreases with temperature

**Table 1.** This is a table summary of the total area its distribution on left (area2) and on right (area1) the isosbestic point (at 1438nm) under the baseline corrected absorbance spectra of Milli-Q water samples of 20-74 °C in the 1300-1600 nm spectral range.

| TABLE1 1 | | | | | | |
|---|---|---|---|---|---|---|
| temp, | area1 | ± 3xSD | area2 | ± | area1/area2 | total |
| 20 | 89.15 | ± 0.26 | 44.18 | ± 0.29 | 2.02 | 133.33 |
| 22 | 88.55 | ± 0.16 | 44.84 | ± 0.18 | 1.97 | 133.39 |
| 24 | 88.24 | ± 0.47 | 45.18 | ± 0.51 | 1.95 | 133.42 |
| 26 | 87.59 | ± 0.37 | 45.88 | ± 0.40 | 1.91 | 133.47 |
| 28 | 87.08 | ± 0.26 | 46.42 | ± 0.28 | 1.88 | 133.50 |
| 30 | 86.50 | ± 0.23 | 47.05 | ± 0.24 | 1.84 | 133.55 |
| 32 | 85.95 | ± 0.15 | 47.63 | ± 0.15 | 1.80 | 133.58 |
| 34 | 85.54 | ± 0.37 | 48.04 | ± 0.39 | 1.78 | 133.58 |
| 36 | 84.78 | ± 0.05 | 48.83 | ± 0.05 | 1.74 | 133.62 |
| 38 | 84.24 | ± 0.02 | 49.39 | ± 0.02 | 1.71 | 133.62 |
| 40 | 83.86 | ± 0.22 | 49.77 | ± 0.22 | 1.69 | 133.62 |
| 42 | 83.33 | ± 0.23 | 50.31 | ± 0.26 | 1.66 | 133.64 |
| 44 | 82.77 | ± 0.34 | 50.87 | ± 0.33 | 1.63 | 133.65 |
| 46 | 82.24 | ± 0.44 | 51.38 | ± 0.47 | 1.60 | 133.62 |
| 48 | 81.61 | ± 0.17 | 52.00 | ± 0.15 | 1.57 | 133.61 |
| 50 | 81.22 | ± 0.50 | 52.39 | ± 0.48 | 1.55 | 133.61 |

| | | | | | |
|---|---|---|---|---|---|
| 52 | 80.60 | ± 0.24 | 52.97 | ± 0.22 | 1.52 | 133.57 |
| 54 | 80.29 | ± 0.36 | 53.26 | ± 0.34 | 1.51 | 133.55 |
| 56 | 79.65 | ± 0.14 | 53.88 | ± 0.15 | 1.48 | 133.53 |
| 58 | 78.98 | ± 0.25 | 54.54 | ± 0.21 | 1.45 | 133.53 |
| 60 | 78.28 | ± 0.47 | 55.15 | ± 0.43 | 1.42 | 133.43 |
| 62 | 78.24 | ± 0.56 | 55.19 | ± 0.53 | 1.42 | 133.44 |
| 64 | 77.68 | ± 0.20 | 55.71 | ± 0.17 | 1.39 | 133.38 |
| 66 | 77.19 | ± 0.33 | 56.14 | ± 0.30 | 1.37 | 133.33 |
| 68 | 76.59 | ± 0.07 | 56.69 | ± 0.06 | 1.35 | 133.28 |
| 70 | 75.92 | ± 0.24 | 57.28 | ± 0.20 | 1.33 | 133.20 |
| 72 | 75.62 | ± 0.63 | 57.56 | ± 0.54 | 1.31 | 133.18 |
| 74 | 75.08 | ± 0.51 | 58.02 | ± 0.42 | 1.29 | 133.11 |

By plotting the values from table 1, it's possible to see that such a system exhibits a van't Hoff behavior, i.e. the constant of equilibrium of the reaction (here the equilibrium between the two components of water) is related to the Gibbs energy variation, $\Delta G$, at the equilibrium by

$$\Delta G_{equil} = \Delta H_{equil} - T\Delta S_{equil} = -RT \ln K_{equil} \tag{1}$$

$$\text{where} \quad \Delta H_{equil} = E_2 - E_1 \quad \text{and} \quad \ln K_{equil} = \ln \frac{A_2(\lambda,T)}{A_1(\lambda,T)} \tag{2}$$

and R is the ideal gas constant R=8,3144 J K$^{-1}$mol$^{-1}$ = 8,617 × 10$^{-5}$ eV K$^{-1}$ atom$^{-1}$.

The van't Hoff plot can be drawn by bisecting the spectrum at the isosbestic point and plotting the natural logarithm of the ratio of the areas above and below that point versus 1/T, see fig. 5. $A_1(\lambda,T)$ is referred to the spectral intensity of the oscillators peaked at lower energy (and larger wavelength) $A_2(\lambda,T)$ is referred to the spectral intensity of the oscillators peaked at higher energy (and shorter wavelength).

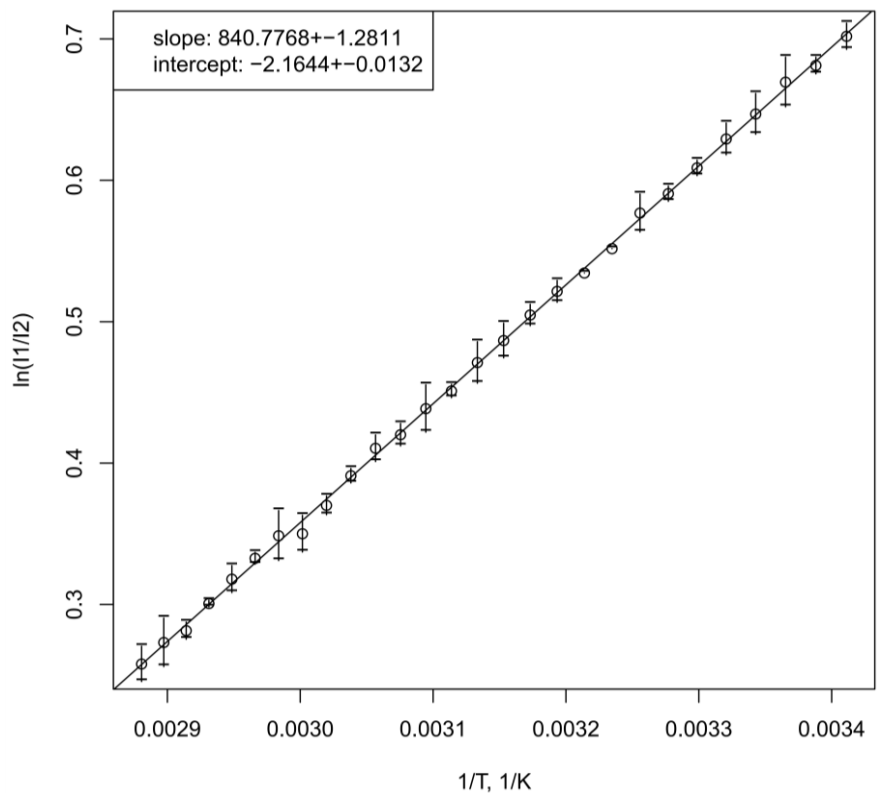

**Figure 6.** The van't Hoff law agreement within the 20-74°C range (corresponding to 293-347 K range)

We obtain a straight line whose slope corresponds to the energy difference between the two vibrational components 1 and 2[1]:

$$-RT \ln \frac{A_2(\lambda,T)}{A_1(\lambda,T)} = E_2 - E_1 - T\Delta S_{equil} \rightarrow \ln \frac{A_1(\lambda,T)}{A_2(\lambda,T)} = \frac{E_2-E_1}{R}\frac{1}{T} - \frac{\Delta S_{equil}}{R} \quad (3)$$

corresponding to the equation of a straight line like $y = mx + c$, where $x = \frac{1}{T}$, the slope is $m = \frac{E_2-E_1}{R}$ and the intercept is $c = -\frac{\Delta S_{equil}}{R}$. Figure 6 shows a sharp linearity thus confirming that the difference of energy of the two phases does not depend on temperature in the range under investigation. The slope of the curve $\Delta E/R$ gives[2]: $\Delta E=6.99\pm0.01$ kJ/mol (or 0.72 eV per molecule) which is a bit smaller than what has been reported in ref. [17]; this value is normally associated to the enthalpy change in the hydrogen-bond rupture [20-22], however, this value is much lower than the value 18,82 kJ/mol (4.5 kcal/mol, or about 0.2 eV) per molecule estimated from the enthalpy of sublimation of ice. The intercept of the straight line is equal to the molar entropy change $\Delta S/R$ between the two populations, here $T\Delta S=5.27\pm0.03$ kJ/mol at T=293 K, about the 75% of the estimated value of $\Delta E$. The positive slope implies that the energy of the population-2 (which increases with T) is higher than the energy of the population-1. The negative intercept ($-\Delta S/R$) shows $\Delta S=S_2-S_1>0$ indicating that the population-2 lives in a more disordered state. This shows the higher entropy of the high-density phase of water and supports the assignment of the population-1 with the coherent phase envisaged by QED [9,11].

The existence of the isosbestic point has been previously interpreted [20-22] as the result of changes in a continuous distribution of H bonding geometries and of the microscopic nature of the fluctuations of the (single state) thermal bath. A single species coupled to a Gaussian bath could satisfy the van't Hoff behaviour equation. However, the demonstration always implies the existence of two species, artificially dividing the spectrum at a cut-off frequency whose physical reason has not been well defined [20].

Another relevant aspect we noticed in the first overtone region is the linear dependence of the logarithm of the ratio ($I_1/I_1+I_2$) on the logarithm of temperature, shown in Fig. 7.

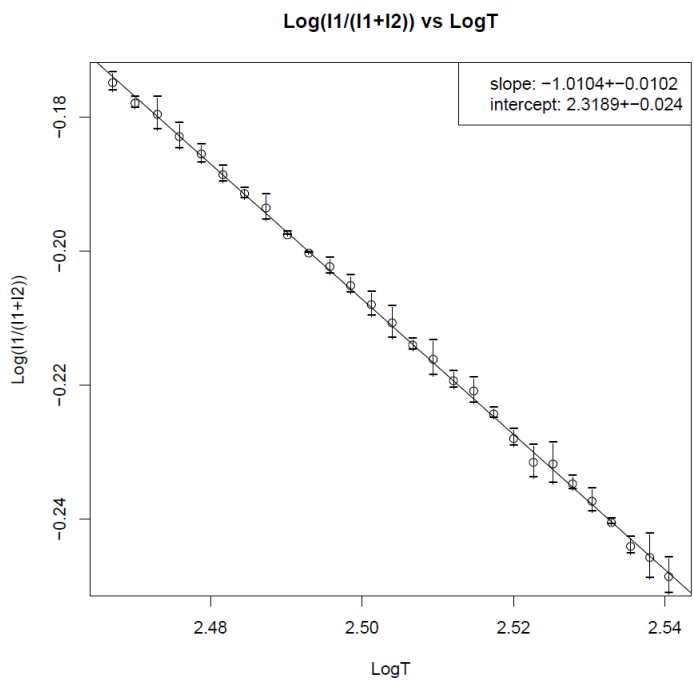

---

[1] Important: $E_1$ and $E_2$ must not to be confused with the energies corresponding to $\lambda_1$ and $\lambda_2$ in the spectra! $E_1$ and $E_2$ are indicated only as the (not defined) terms of $\Delta E$, that is: the energy difference between the fundamental states of the population-1 and of the population-2. $\Delta E$ is calculated from the slope of the van't Hoff line that, along with the intercept $\Delta S/R$, are the values we obtain by plotting the data, as explained above and shown in fig. 5.

[2] The errors reported come from the fit.

**Figure 7.** Log($I_1/I_1+I_2$) vs Log(T) in the 20°-74° C range. The linear behavior highlights a scale-free (power) law, witnessing that relevant collective dynamics underlies the system response to perturbations (like temperature increasing).

Such a behavior has been already observed by one of us in the IR range for aqueous solutions at different ions' concentrations. The independence of this log-log linear trend of the kind of solute [49] suggests this feature is typical of a long-range ordering in the solvent (i.e.: liquid water).

## 4. Discussion

NIR spectroscopy already looked at water in condensed phases (liquid, ice, confined states, etc.) as a "matrix" where the properties of the "whole" can well reflect the status of the components inside it. Unexpected information has been obtained about the presence of really tiny perturbations (like very low concentrations of ions) [23-25]. Moreover, it has been shown that, by detecting the changes of the spectral features in the system and performing adequate meta-data analysis, it's possible to trace back the kind and the magnitude of the perturbations (temperature, electrolytes, pH changes, solute concentration, aging, effect of light measurements, etc.) so revealing fine details about the "history" of the sample by slight changes in the acquired spectra and by matching them each other [26-29].

Vibrational energies of intramolecular OH bonds absorb IR radiation. In NIR these vibrations can be studied in the first overtone range. In the third table of ref. [1], among the others, an absorption peak at about 1470 nm is reported. We identified, by second derivative deconvolution, two peaks: at 1412 nm and 1462 nm. The difference of the second one with respect to the value reported in ref. [1] can be attributed both to the precision of the instrument (its sensitivity is 8 nm in our spectral range) and to the experimental layout: we work in transmittance and in quite confined cuvettes, where water is shaped in form of a thin layer included between two moderate-good hydrophilic surfaces. The two peaks at 1412 nm and at 1462 nm corresponds to the two populations/states. The existence of two populations (whose OH-bond vibrational states lay below and above an isosbestic point) has been previously explained [30] by the "mixture-model" requiring an *ad hoc* cut-off frequency at the isosbestic point. Moreover, the energy difference between the two populations is independent of temperature which excludes the hypothesis that their difference is only due to different geometries of the H-bond network (which would be greatly affected by temperature). Actually, the van't Hoff's plot defines the equilibrium constant of the chemical-physical reaction and provides the Gibbs' free energy of the transition independent on temperature. The ΔH obtained from the slope of van't Hoff line fits quite well with the energy gap ($E_g$) defined in QED theory of water as a two-fluids liquid [13].

The existence of two different populations (having to different energy states) within liquid water, made up of molecules having different kind of mutual interactions, has been already predicted in the framework of QED [10,11,13]. In this picture, liquid water is made up of two populations whose relative abundance is function of the temperature. It has been shown that the interaction among particles is mediated by electromagnetic fields emerging from their collective quantum oscillations [9-14,31]. These fields induce long-range forces, which are responsible for the existence of the condensed state provided that the density is above a critical density and the temperature is lower than a critical temperature.

We measured ΔH≈ 0.07 eV however, such a number is smaller than the value obtained for a similar measure in IR range (1000-4000 cm-1), 0.12 eV as reported in ref. [17]. Such a difference can be attributed to differences in the experimental layout: in ref. [17] total reflectance technique (ATR-IR) has been used where a water droplet is placed on the IR-transparent surface. On the contrary, we operated in transmittance: the volume of the liquid was confined into quartz cuvettes with high surface/volume ratio (see figure 2). Silicon dioxide cuvette walls are a moderate-to-medium hydrophilic material [16]. It is well known that liquid water close to hydrophilic surfaces, or in confined situations has a very peculiar behavior not yet completely understood and it is well accepted that the interaction with the container may affect the long-range arrangement [32-37]. Thus, we make the hypothesis that the experimental set-up might affect the result in case of high surface volume ratio geometries. Comparing qualitatively the ratios between $I_1$ and $I_2$ with QED calculations [35] in the analyzed range of temperature, we estimate that about 10-20% of the light path travelled by the probing beam is made of interfacial water, enriched in one of the two phases (the coherent one, according to QED), with respect to the bulk.

Moreover, it's noteworthy that coherence on electronic oscillations has been demonstrated to be unstable in purely three-dimensional (3D) systems [50], and it has been referred to ideally pure "bulk" water too, expressing that the coherence domains are unstable. In [50] indeed it has been pointed out that the QED theory of water [9-11,13] has been built by neglecting interactions between quasi-free coherent electrons associated to each coherent domain in the bulk and that the stability of the latter is possible therefore only if electron density is not too high (upper limit) even though a minimum threshold must be overcome (lower limit) in order to onset a sufficient coupling between electromagnetic field and the $N$ pushed-in-oscillation charges (i.e.: the Lamb-shift term, proportional to $N\sqrt{N}$) [9,11]. The possibility of stable coherence domains has therefore been deemed possible only in presence of whatever kinds of interfaces (walls, molecular chains, nanobubbles, nanoparticles or even some kinds of solutes), [50] able to create 2D configurations [9,11]. Actually, the theory foresees the existence of quasi-free electrons only at the surface defining the border between coherent and non-coherent phase. Each coherence domain, acts as a self-produced cavity. This means that each domain has a sharp interface (few Å thick [9,11]) within decay profile of the field. At the moment, the real topology of coherence domains is still unknown. However, the restrict range considered in [50] is obeyed in explicitly interfacial situations, like Exclusion Zone Water [32-34], water sheaths around biomolecules and membranes, etc.

Another important consideration to be made deals with liquid water condensation: the existence of several excited levels, working as possible candidates for other coherent states (and respective phases), has been suggested as the origin of the existence of several kinds of ice [13]. Experimental conditions may, in point of fact, affect the oscillator strengths of the individual levels thus selecting different oscillations between the ground state and this level. Further analysis on this important topic are needed in order to correctly match the measured and calculated energy gaps.

The importance of collective behaviors in long range organization of water molecules is also suggested by the log-log plot shown in fig. 7 which relates the logarithm of the spectral intensity to the logarithm of the applied perturbation (T). Scale-free models ($y = x^n$) have received a great attention recently because they describe the auto-organization in complex systems. Whenever a complex system (ecosystem, neuron net, cellular metabolism [38,39]) shows a log-log, scale-free behavior we are entitled to predict a complex long-range organization. Such a behavior has been also detected in FTIR spectroscopy of water solutions of alkali halides on changing the concentration of the solutes pointing towards a very general property of water restructuring under perturbations [49]. Moreover, Vitiello et al. [40-44] have well shown that a functional representation of self-similarity is mathematically isomorphic to quantum coherent states. Quantum coherence, beyond dissipation [39,45], thus appears to be the prime candidate to underlie the ubiquitous recurrence of self-similarity (and fractals) in Nature both in physics and in biology [46-48].

## 5. Conclusions

We showed that a very simple system (Milli-Q water) perturbed by the temperature shows features not explainable within a simple molecular-dynamics approach. The OH first stretching-mode overtone (1300-1600 nm range) is featured by the presence of a clear isosbestic point that reveals the existence of two populations of molecules in the liquid. The non-exact nature of the isosbestic point provides an additional hint on the linear dependence of the relative abundance of the two populations on the temperature in the analysed range of temperature. As shown by the linearity of the van't Hoff plot, their fundamental states have an energy difference independent of temperature, so the "mixture model", requiring an *ad hoc* cut-off frequency at the isosbestic point, is not experimentally supported. The slope of the van't Hoff line provides an energy difference which matches (within the errors) the *energy gap*, $E_g$, calculated in the QED theory of liquid water.

The mismatch between our result and that obtained with ATR-IR spectroscopy shed a light on the role of the water-surface contact whenever a quantitative evaluation of the spectra is needed. A more accurate investigation about this point should be carried on in the next step of this research.

The scale-free behavior emerging from the analysis of the effect of the temperature, suggests an underlying coherent dynamic. The two-fluids water cannot be fully described within an electrostatic (classical) picture but requires a deeper quantum approach [9-14]. We hope this paper can contribute to the stimulating discussion about the water structure. We envisage that studies on the liquid water will be considered in a next future in the realm of the emerging *science of complexity* [38,47,48].

**Author Contributions:** First conceptualization, P.R., A.DN. and R.T.; methodology, R.T., Z.K.; software, Z.K.; investigation, Z.K.; resources, R.T.; data curation, Z.K.; writing—original draft preparation, P.R., A.DN; writing—review and editing, P.R., Z.K.; supervision, R.T., A.DN.; project administration, R.T..

**Funding:** This research received no external funding.

**Acknowledgments:** The authors heartfeltly thanks give Giuseppe Vitiello for precious and useful discussions about fractal and squeezed coherent states in QED and Livio Giuliani for their help in laying down a well-focused writing and for fruitful discussions about the physical meaning of the phenomena. Perennial estimations and tributes addressed to Emilio Del Giudice and Giuliano Preparata (no more among us), our wise, unmatchable masters and inspirers. The author ZK gratefully acknowledges the support by the János Bolyai Research Scholarship of the Hungarian Academy of Sciences, Hungary and by the Hungarian Government through project No. EFOP-3.6.3-VEKOP-16-2017-00005. The author PR send his heartfelt thanksgiving to all Shigeoka's Family, to Mitsue Oshima and to all the Yunosato facility (Hashimoto, Japan) for their precious support and careful hospitality.

**Conflicts of Interest:** The authors declare no conflict of interest.

## Appendix A

M. Greger et al. [19] in 2013, have developed a framework which allows one to extract information about a correlated system by analysing the sharpness of an approximate isosbestic point in a physical quantity *f(x; p)*. Isosbestic behaviour is usually observed only in a certain parameter range around some particular value $p_0$ and can be expected to break down away from $p_0$. In our experiment the parameter *p* is embodied by temperature, *T*, and the *x* variable is the wavelength, *λ*.

Crossing points among the spectra are, in general, located on a curve *λ\*(T)*, such as

$$\left.\frac{\partial I(\lambda,T)}{\partial T}\right|_{\lambda*(T)} = 0 \qquad \text{A1).}$$

In the case all the spectra intersect in a single point, $\lambda^*$, it means that $\lambda^*$ does not depends on T. We have observed an approximate isosbestic point, i.e. a crossing region 8 nm wide. Such an approximate isosbestic point corresponds to a weak dependence on T. In order to evaluate this dependence we can expand our function I(λ,T) around T=$T_0$:

$$I(\lambda, T) = I(\lambda, T_0) + (T - T_0)F_n(\lambda, T_0) + \mathcal{O}[(T - T_0)^{n+1}] \qquad \text{A2)}$$

where $F_n(\lambda, T) = \frac{\partial^n I(\lambda,T)}{\partial T^n}$ and $O[(T-T_0)^{n+1}]$ represents the series components of higher order than $n^{th}$. So that one gets the approximated function as

$$\tilde{I}(\lambda, T) = I(\lambda, T_0) + (T - T_0)F_n(\lambda, T_0) \qquad \text{A3)}$$

which can yield an exact isosbestic point-like behaviour when the term $O[(T-T_0)^{n+1}]$, responsible for deviations from the exact isosbestic point, is omitted.

Therefore, in order to isolate the dependence of wavelength on temperature, we expressed the spectral intensity (the absorbance) of each spectrum as a power series up to the first order, made of two terms: the first independent on T, the second dependent on T:

$$I(\lambda(T), T) \rightarrow \tilde{I}(\lambda, T) = I(\lambda)|_{T_i} + (T_i) \cdot I_1(\lambda)|_{T_i} + \mathcal{O}[T_i^2] \qquad \text{A4),}$$

where *i*=1,2,3,4,5,6 (see in fig. 1A the six chosen temperatures: 20°C, 30°C, 40°C, 50°C, 60°C, 70°C) and

$$I_1(\lambda) \simeq \frac{I(\lambda, T_b) - I(\lambda, T_a)}{T_b - T_a} \qquad \text{A5).}$$

To determine $I_1(\lambda)$ we chose $T_a$=30°C and $T_b$=70°C. In Fig. A1 we show the comparison between the absorbance I(λ,T) at six different temperatures compared with the approximated absorbance $\tilde{I}(\lambda, T)$ The linear temperature dependence of $\tilde{I}(\lambda, T)$ explains the sharpness of the isosbestic point and support the hypothesis that the absorbance is proportional to a function A(λ, 0)+ TA$_1$(λ), being A(λ) the deformation of the spectrum in the range of temperature analysed.

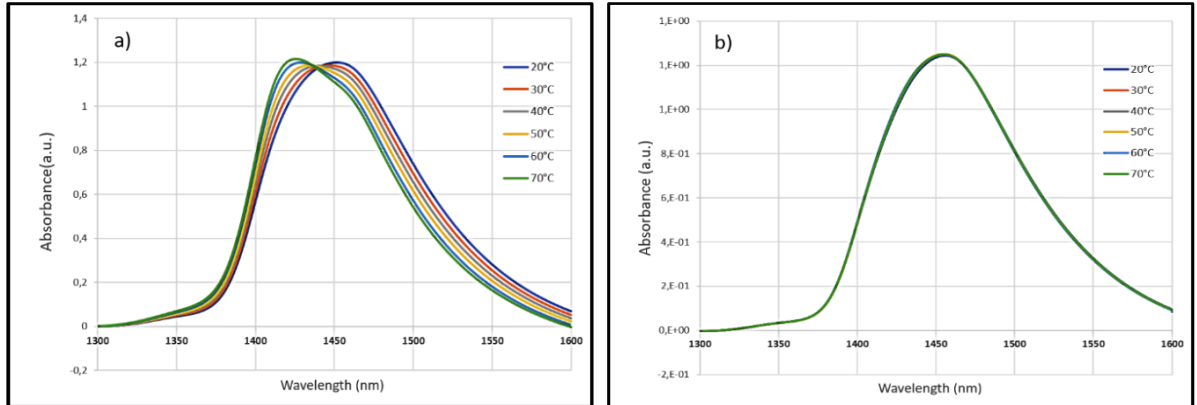

**Figure A1. a)** The spectra (averaged over 3 scans at each T) acquired at six different temperatures (20-70°C) are shown. The isosbestic behaviour is clear, but it is not perfectly sharp: it spans in an 8 nm wide region. **b)** The same spectra after the perturbative procedure explained in the appendix text confirms that the dependence of the crossing point, λ*(T), is linear in T.

Based on the behavior of the spectra we can thus identify three different cases: a) no crossing point means that the effect of the temperature is the same all over the frequency range, i.e. is the same for all the absorbing molecules; b) the existence of an exact crossing point implies the existence of two population of molecules in the sample whose relative abundance is regulated by the running parameter; c) an approximate isosbestic point also gives information about the nature of the dependence of the spectra by the parameter.

In case of liquid water, the temperature linear dependence of equation A2) supports the ansatz that the relative abundance of the two populations in water is linearly regulated by T, in a small range of temperature around T$_0$, at room conditions [11].

**Appendix B**

In this brief section we give some graphical details about the procedure adopted for the baseline correction.

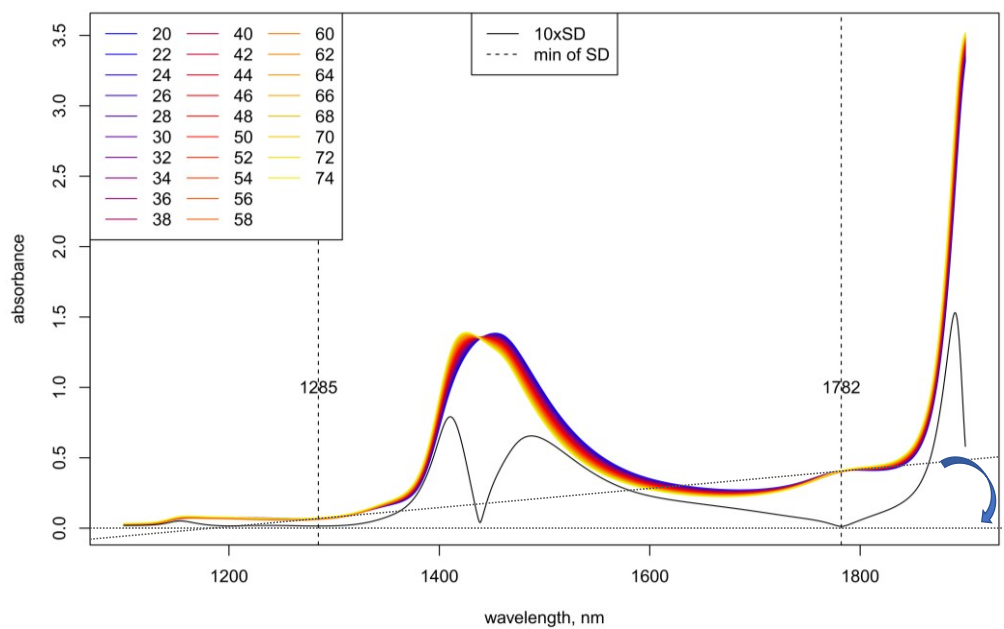

**Figure B1**. Raw absorbance spectra of the Milli-Q water samples of 20-74 °C in the spectral range of 1100-1900 nm and presentation of the calculation steps of the baseline correction. (1) calculation of the standard deviation (SD)of the absorbance values (continuous line), (2) determination of the minimums of the SD (dashed line, 1285 and 1782 nm), (3) fitting linear models at 1285 and 1782 nm in all the single spectrum (dotted line) which can be used for the rotation to zero offset and zero slope (Figure B2).

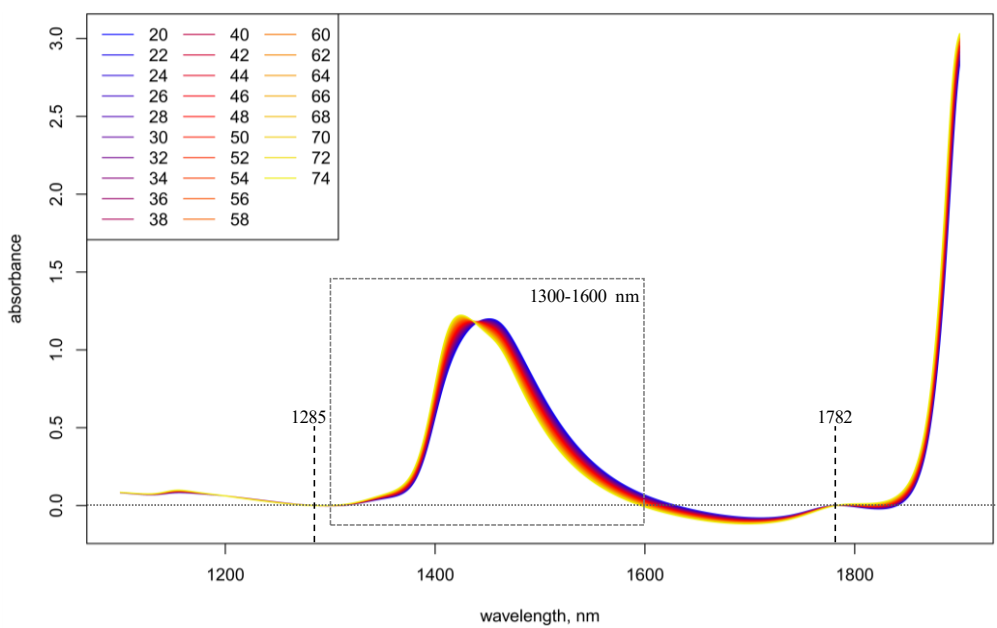

**Figure B2.** Baseline corrected absorbance spectra of the Milli-Q water samples of 20-74 °C in the spectral range of 1100-1900 nm showing absorbance values zero at 1285 and 1782 nm (as a result of the transformation) highlighting (framed with dashed lines) the first overtone range of water (1300-1600 nm) used for further calculations.